# Analytical Form of Forces in Hydrophobic Collapse

J. Chakrabarti[1] and Suman Dutta

Department of Chemical, Biological and Macromolecular Sciences,

[1]Also at the Thematic Unit of Excellence in Computational Material Science,

S N Bose National Centre for Basic Sciences, Block-JD, Sector-III, Kolkata 700 098, India

Abstract

We calculate analytically the forces between two solvophobic solutes, considering a model system. We show that the effective interaction forces between two solvophobic solutes, mediated by the solvent, is attractive for short ranges, which decreases linearly with surface-to-surface separation $s$ between the solutes and repulsive in the long range falling off as $1/s^4$. The attraction originates from the unbalanced Laplace force at the liquid-gas interface, generated by the repulsive interaction with the solvent particles, around the solutes at small $s$. The long range part arises due to unbalanced osmotic pressure. We illustrate the calculations for the Lennard-Jones solvent. We discuss the general implication of our results in the context of hydrophobic collapse.

Hydrophobic species in aqueous solution tend to form aggregation in order to avoid water contact. Although hydrophobicity mediated collapse lies at the heart of a host of chemical and physical phenomena [1], the nature of the forces associated with the collapse is yet poorly understood.

There have been numerous efforts to uncover the hydration behaviour of hydrophobic species in an aqueous medium. The effects of the aqueous medium depend on the size of the hydrophobic species [2]. For instance, a large hydrophobic surface disrupts the hydrogen bond network in water. The hydration of large hydrophobic species is thus enthalpy dominated. The water molecules avoid the vicinity of the hydrophobic surface, creating a low density interface of partially ordered water molecules [1], analogous to gas-liquid interface. However, small hydrophobic molecules do not disrupt the hydrogen bond network, indicating an entropy dominated hydration. The current theoretical understanding of hydration of hydrophobic species in water exploits the Gaussian nature of equilibrium density fluctuations over the uniform system [3].

The collapse of hydrophobic species in an aqueous solution is driven by attractive effective forces [4] between the hydrophobic species mediated by water, as evidenced by different measurements [5]. Computer simulation studies [6] indicate that the interaction between a large hydropohobic solute with water can be modeled in terms of solute-solvent repulsion, generally termed as solvophobic interaction. Numerical simulations on model systems reveal short ranged effective attraction and long ranged repulsion between two solvophobic solutes [7,8]. Such interaction forces have been invoked to explain structure and dynamics of many systems, including protein solutions [9].

Here we develop a simple analytical theory for the solvent mediated effective interaction forces between two large solvophobic solutes at a separation s in a solution. We show that the effective interaction forces between two large spherical solvophobic particles are attractive for short ranges, decreasing linearly with *s* and repulsive in the long range falling off as $1/s^4$. Such simple forms of the force laws associated with collapse of solvophobic particles may be useful to understand their structural and dynamical behavior in solution.

Let us consider for simplicity two model solvophobic solutes in a model solvent with the Lennard-Jones (LJ) interaction potential $V(r) = 4\varepsilon\left[\left(\frac{r}{\sigma}\right)^{12} - \left(\frac{r}{\sigma}\right)^{6}\right]$ between two solvent particles at separation *r*, $\varepsilon$ and $\sigma$ being the interaction parameters. The large density inhomogeneity of the gas-liquid interface is taken into account via the classical density functional theory (DFT) [10]. This gives the radius $\Lambda_0$ of the gas bubble surrounding the solute of radius *R*. According to the classical DFT the free energy cost of creating local density in-homogeneity $\delta\rho(\vec{r}) = \rho(\vec{r}) - \rho_0$ at a point $\vec{r}$ over the uniform density $\rho_0$ in the presence of an external potential $V(\vec{r})$, is given by[10]:

$$\frac{F}{k_B T} = \int d\vec{r}\rho(\vec{r})\ln\left(\frac{\rho(\vec{r})}{\rho_0}\right) - \frac{1}{2}\int d\vec{r}d\vec{r}'\, c^{(2)}(\vec{r}-\vec{r}')\delta\rho(\vec{r})\delta\rho(\vec{r}') + \int d\vec{r}V(\vec{r})\delta\rho(\vec{r}). \qquad (1)$$

Here $k_B$ is the Boltzmann constant and $T$ the temperature. The first term in Eq.(1) corresponds to the entropic contribution. The second term corresponds to the correlation term arising due to the inter-particle interaction in the liquid, $c^{(2)}(\vec{r}-\vec{r}')$ being the liquid direct correlation function [10]. The solute may be thought of as generating an external repulsive potential $V(\vec{r})$ in the liquid medium of density $\rho_l$, which stabilizes a gas bubble of density $\rho_g$ around the solute. We assume that the gas-liquid interface is sharp: the medium density is $\rho_g$ the density of the gas in the bubble and is $\rho_l$ just outside the bubble. The equilibrium bubble radius $\Lambda_0$ can be estimated by minimizing the free energy cost of creating a bubble of radius $\Lambda$ around the solute at $\vec{X}$ in the liquid solvent. Here $\delta\rho(\vec{r}) = \rho_g - \rho_l$ over all the points within the shell bound by R and $\Lambda$, and otherwise, the density difference is zero. The free energy difference between the bubble and the liquid within the cavity, as given by the DFT:

$$\frac{F}{k_B T} = V_0 \rho_g \ln[\rho_g/\rho_l] - \frac{V_0}{2}(\rho_g - \rho_l)^2 \int_0^\infty d^3 r_{ij} c^{(2)}(r_{ij}) + (\rho_l - \rho_g)\int_R^\Lambda V_{bs}(\vec{r}-\vec{X})d^3r + 4\pi\gamma\Lambda^2 \quad (2)$$

where $V_0 = 4\pi[\Lambda^3 - R^3]/3$. In addition to the terms in Eq.1, the last term in eq. (2) is the free energy cost of creating the gas-liquid interface at the bubble surface, $\gamma$ being the surface tension.

The minimization of $\frac{F}{k_B T}$ with respect to $\Lambda$ yields an algebraic equation to get an equilibrium value of the bubble radius, $\Lambda_0$:

$$4\pi\Lambda_0^2\left[\rho_g \ln\left(\frac{\rho_g}{\rho_l}\right) - \frac{c_0}{2}(\rho_g - \rho_l)^2 + (\rho_l - \rho_g)V_{bs}(\Lambda_0)\right] + 8\pi\gamma\Lambda_0 = 0, \quad (3)$$

using the Leibniz rule. One can get an approximate analytical solution of Eq. (3) for model solvophobic repulsive interaction, $V_{bs}(\vec{r}-\vec{X}) = \varepsilon_{bs}(\sigma/|\vec{r}-\vec{X}|)^n$, $\varepsilon_{bs}$ being the strength of the interaction and $n$ the exponent describing the steepness of the interaction, for $n>>1$. In such case, $\frac{\Lambda_0}{R} \approx \left[\frac{\tilde{\varepsilon}_{bs}}{(\delta\rho|c_0|+\gamma R^2)}\right]^{1/n}$.

Here $\tilde{\varepsilon}_{bs} = \varepsilon_{bs}/k_B T$ and $c_0 = \int d^3 r c^{(2)}(r)$, the zero wave-vector Fourier component of $c(r)$ which is related to the compressibility of the liquid solvent, $\frac{\rho_l \chi_T}{k_B T} = \frac{1}{1-\rho_l c_0}$ [10]. If we assume that $\rho_l >> \rho_g$, so that $\delta\rho \approx \rho_l$ and

$\frac{\Lambda_0}{R} \sim \left[\frac{\tilde{\varepsilon}_{bs}}{(1-k_B T/\chi_T)}\right]^{1/n}$ with low $\gamma$ at the interface. In two dimensions, our estimation of $\frac{\Lambda_0}{R}(\approx 1.3)$ is very good for the system simulated earlier [7] in subcritical conditions with $n=12$. In three dimensions for similar system parameters, $\frac{\Lambda_0}{R} \approx 1.27$ for the isotherm $k_B T/\varepsilon = 1.25$, using (dimensionless) $\gamma = 0.7$.

A solute particle, surrounded by a vapour of radius $\Lambda_0$, will be subject to the Laplace pressure difference $\delta P_0 = 2\gamma/\Lambda_0$ at the liquid-gas interface of the bubble due to the interfacial tension. The pressure is larger inside the bubble than that on its outside, the excess pressure being balanced by the Laplace pressure. Let us now consider a pair of solvophobic solute particles. If two solvophobic macromolecules each of radius $R$, surrounded by the bubbles of radius $\Lambda_0$, are brought to a close separation with the surface-to-surface separation, $s \leq 2\Lambda_0$, the solvent particles would be depleted from the overlapping region due to the solvophobic interactions. Consequently, the interfacial tension driven Laplace pressure would vanish in the overlapping region due to expulsion of the solvent particles, while the other parts of the solutes are subject to the Laplace pressure. The anisotropic distribution of the solvent particles, shown in Fig.1, thus results in a local pressure gradient, leading to an attractive force between the two solute particles.

The magnitude of the attractive forces on the solute particles can be estimated by integrating the pressure over the particle surfaces. Let us consider the geometry shown schematically in Fig. 1 and define $z = 2R + s$. The overlapping region is given by the polar angle $\theta$ in the range $-\cos^{-1}(z/2\Lambda_0) \leq \theta \leq \cos^{-1}(z/2\Lambda_0)$, assuming the azimuthal symmetry. The area of the gas bubble, facing the overlapping region, $A = \pi z^2 \tan^2[\cos^{-1}(x)] = 4\pi R^2 \Lambda_0^2 (1-x^2)$ with $x = z/2\Lambda_0$, over which the Laplace pressure is zero. The magnitude of the net unbalanced force on the bubble is then given by: $F = -\delta P_0 4\pi R^2 \Lambda_0^2 (1-x^2)$. The negative sign indicates the decrease in pressure due to the disappearing of the liquid-gas interface in the overlapping region. Since the same pressure will act on the sphere so that the attractive force on the sphere is given by, $F_A = -\left(\frac{\Lambda_0}{R}\right)^2 F$. Inserting the definition of $x$ and $z$, we get $F_A = -\alpha\left[\left(1-\left(\frac{\Lambda_0}{R}\right)^2\right) - \frac{Rs}{\Lambda_0} - \frac{s^2}{4\Lambda_0^2}\right]$, where $\alpha = 4\pi\Lambda_0\gamma/R$. This indicates that the attractive force is maximum at contact between two solutes (*s=0*), then decreases linearly with *s* for small *s* and finally vanishes when two bubbles just touch each other ($s = 2(\Lambda_0 - R)$). This sort of attractive forces between solvophobic solutes have been reported in earlier studies [7,8]. The solvophobicity mediated attraction strength given by the parameter $\alpha$ depends both on the bulk solvent property via $c_0$ and the interfacial tension $\gamma$.

Let us also consider the asymptotic form of interaction forces between solvophobic solutes for large $s$ ($\leq 2\Lambda_0$). When the pair of solvophobic particles is pushed apart, the confined region between the gas bubbles is filled up with the solvent. However, the solvent density in this region will be different than that in the bulk. In this case the Laplace pressure will be balanced as in the case of single solvophobic particle. However, there will be unbalanced osmotic pressure. The intervening region can be thought of as a cylindrical capillary with height $\Lambda_0$ and radius $z/2$. The solvent density difference in this region with respect to the bulk, from the condition of equality of the chemical potential $\mu$ with that in the bulk, $\delta\rho = \dfrac{\rho_l \mu}{V k_B T (1 - \rho_l c_0)}$ where $V \approx s^2 \Lambda_0$. Since $c_0 < 0$ for a liquid, the density in the confined capillary is larger than the bulk. Moreover, the density difference vanishes for very large $z$. The excess pressure, $\delta P \sim \delta\rho k_B T$, due to the enhanced density would push the two spheres apart, resulting in repulsion. The repulsive force on the bubble, $F' = \delta P \Lambda_0^3 / s^2 (2\Lambda_0 + s)^2$, obtained by integrating the pressure over the area of the bubble exposed to the high density region, the limit of the integration being $\pm \tan^{-1}\left(\dfrac{\Lambda_0}{2\Lambda_0 + s}\right)$. The repulsive force on the sphere is then given by $F_R = \left(\dfrac{\Lambda_0}{R}\right)^2 F'$. For large values of $s$ such that $s \gg 2\Lambda_0$, $F_R \sim s^{-4}$.

The case of a LJ fluid in three dimensions is illustrated. We calculate the DCF, $c^{(2)}(r)$ using the standard prescription of the liquid state theories [10]: The short-ranged part of the DCF is taken to be of the Percus-Yevick form for hard sphere representing the steep repulsive part of the LJ interaction. The long ranged part is treated in the mean field approximation, replacing the correlation function by the attractive tail of the LJ potential with an opposite sign. We restrict only to the liquid region bound by the triple point temperature $T_R$ and the liquid-gas critical point temperature $T_c$ in the LJ phase diagram [10]. The plots of $\rho_l c_0$ versus $\rho_l$ are shown in Fig. 2(a) for various $T$. The increase in $\rho_l c_0$ with $\rho_l$ for a given $T$, and that with decreasing $T$ for a given $\rho_l$ indicate increasing incompressibility of the liquid.

Let us concentrate on the $T$ dependence. We denote the hard sphere part of $\rho_l c_0$ by $c_0^{PY}$. In the low temperature regime close to $T_R$,

$$\frac{\Lambda_0}{R} \approx \left(\frac{4\pi e}{3}\right)^{1/n}\left[1+\frac{3|c_0^{PY}|k_B T}{4\pi \varepsilon n}\right]$$

to a leading order for large $n$, where $\varepsilon_{bs} = e\varepsilon$. Fig.2 (b) shows $\frac{\Lambda_0}{R}$ as a function of $\frac{k_B T}{\varepsilon}$ for a given $\rho_l$. Experimental data shows that $\gamma$ of a liquid decreases with increasing temperature [11]. This indicates that $\alpha$ has a maximum as a function of temperature, as shown in Fig. 2(c), indicating a maximum of the solvophobocity mediated attraction near $T_R$. Such maximum has been reported by small angle neutron scattering data of tertiary butanol clustering in water [12]. We also examine now the behavior of $\alpha$ in the vicinity of $T_c$. In this regime, $\gamma \sim (T_c - T)$ and $\chi_T \sim (T_c - T)^{-1.33}$ [13], so that $\alpha \sim (T_c - T)^{-0.33}$. The solvophobicity mediated attraction increases with temperature close to the critical temperature, having a divergence at $T_c$. The behaviour of the solvophobic interaction is highly non-monotonic as a function of $T$ with a maximum close to $T_R$ and then divergence near $T_c$.

Short ranged attractive and long ranged repulsive forces have been proposed to explain cluster formation in lysozyme solutions [9]. The short-ranged attractive forces are taken to be generated by the depletion effects [4]. The depletion attraction results from the osmotic pressure difference in the solvent for small separations between the solute particles where the intervening gap is too small to accommodate the solvent particles. However, the surface of lysozyme, as revealed by the crystal structure reported in protein data bank entry 1LZ1, has a large number of hydrophobic pockets. Our calculations suggest that the effective interactions in lysozyme solutions may be of hydrophobic origin. There are several other protein systems known to undergo self-aggregation via cluster formation under hydrophobic forces. The amyloid beta 42 protein fibrils are pathogenic in neuro-degenerative diseases as in the Alzheimer disease [14,15]. Immunoglobulin clusters undergo renal deposit [16]. Chymotripsin aggregates exhibit loss in the enzymatic activity [17]. Self-aggregated biomolecular structures are often pathological: The solvophobicity mediated effective interactions in the present paper, should be applicable to understand cluster formation in all these systems. The qualitative difference between the attractive forces mediated due to solvophobicity and those due to depletion is quite evident. In contrast to the attraction between solvophobic surfaces, the attractive forces due to depletion are due to geometrical constraint. As a result the solvophobicity mediated attraction shows nontrivial temperature dependence, unlike the depletion attraction. There is one important issue in applying the present analysis to realistic systems like proteins where the hydrophobic species often deviates from spherical shape. Although the analysis based on the imbalance of the Laplace pressure will remain valid, the Laplace pressure, governed by the local curvature of the gas-liquid interface, would lead to anisotropic spring constant.

In conclusion we have derived analytical forms of the forces operating between two solvophobic solute particles in a solvent. We show that for small separation, $s$ between large solvophobic particles, the unbalancing Laplace force due to surface tension $\gamma$ at the gas-liquid interface gives rise to attractive force which decreases linearly with $s$. For large separations between the solvophobic particles, the osmotic pressure difference leads to a repulsive force with $1/s^4$ dependence. Such forces have been widely studied as models for dynamical arrest in cluster forming liquids and gels [18-20]. Our calculations suggest that the hydrophobic systems can exhibit analogous cluster formation and interesting dynamical behavior which may underlie rich variety of physical and chemical phenomena exhibited by hydrophobic species in aqueous phase. Detailed exploration of such relationships may be worth investigation in future. Such studies would be relevant in important areas of applications including molecular assemblies, protein crystallization, designing pharmaceutical drugs [21] to name only a few.

**Acknowledgements**
We acknowledge Samapan Sikdar for his help in analyzing the PDB file.

**Figure Captions:**

**Fig. 1:** Schematic of the hydrophobic attraction between two solutes (large filled circles). Due to symmetry about the axis joining the centres of the two solute particles, it is enough to consider the two dimensional projection. The solvent particles (open circles) are expelled from the overlapping (shaded) region, bound by two dashed cicles. The pressure gradient is marked by the arrows: The solid arrows indicate higher pressure due to Laplace pressure and the dotted arrows indicate lower pressure due to absence of the Laplace pressure following the expulsion of the solvent particles. Different geometrical quantities are marked in the figure.

**Fig. 2:** Data for LJ fluid: (a) The dependence of $\rho_l c_0$ with $\rho_l \sigma^3$ for $\frac{k_B T}{\varepsilon}$ =1.2(dotted line) ,1.0(solid line) & 0.7 (dot-dashed line). (b) $\frac{\Lambda_0}{R}$ as a function of $\frac{k_B T}{\varepsilon}$ for $\rho_l \sigma^3$ =0.75. (c) $\alpha \sigma^2$ as a function of $\frac{k_B T}{\varepsilon}$ for n=4, $\rho_l \sigma^3$ =0.75.

**Figures:**

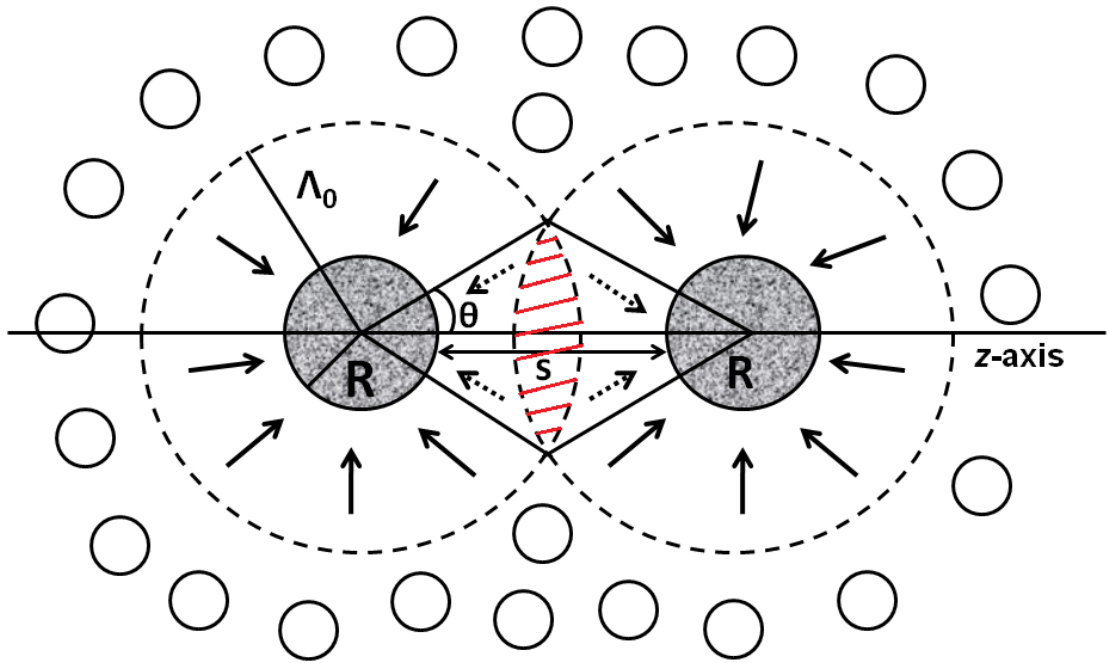

Fig.1: Chakrabarti and Dutta

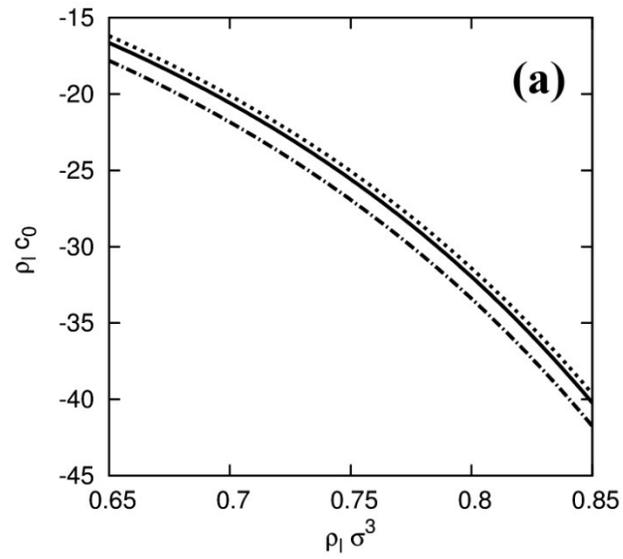

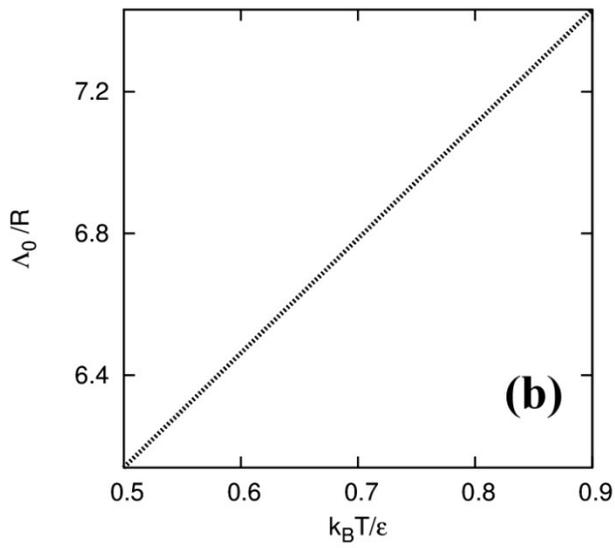

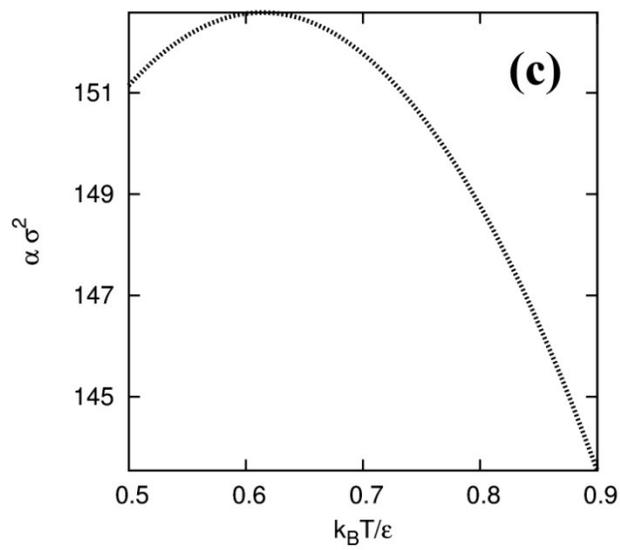

Fig. 2 : Chakrabarti and Dutta